\documentclass[fleqn,usenatbib]{mnras}

\usepackage{newtxtext,newtxmath}
\usepackage{amsmath}
\usepackage{xcolor}

\usepackage[T1]{fontenc}

\DeclareRobustCommand{\VAN}[3]{#2}
\let\VANthebibliography\thebibliography
\def\thebibliography{\DeclareRobustCommand{\VAN}[3]{##3}\VANthebibliography}

\usepackage{graphicx}	
\usepackage{amsmath}	

\title[Interpreting Ly$\alpha$ emission in the EoR]{Ly$\alpha$ with SPICE: Interpreting Ly$\alpha$ emission at $z>5$}

\author[Bhagwat et al.]{
Aniket Bhagwat$^{1}$\thanks{E-mail: abhagwat@mpa-garching.mpg.de},
Lorenzo Napolitano$^{2,3}$,
Laura Pentericci$^{2}$,
Benedetta Ciardi$^{1}$ \&
Tiago Costa$^{4}$
\\
$^1$Max Planck Institut f\"{u}r Astrophysik, Karl Schwarzschild Stra{\ss}e 1, D-85741 Garching, Germany\\
$^2$INAF – Osservatorio Astronomico di Roma, via Frascati 33, 00078, Monteporzio Catone, Italy\\
$^3$Dipartimento di Fisica, Università di Roma Sapienza, Città Universitaria di Roma - Sapienza, Piazzale Aldo Moro, 2, 00185, Roma, Italy \\
$^4$School of Mathematics, Statistics and Physics, Newcastle University, UK\\
}

\date{Accepted XXX. Received YYY; in original form ZZZ}

\pubyear{2015}

\begin{document}
\label{firstpage}
\pagerange{\pageref{firstpage}--\pageref{lastpage}}
 \maketitle

\begin{abstract}
Ly$\alpha$ emission is key to understanding the process of cosmic reionisation.  {\tt JWST} is finally enabling us to measure Ly$\alpha$ emission deep into the epoch of reionisation for an increasing number of galaxies. However, discrepancies between measurements of Ly$\alpha$ equivalent widths (EW$_0$) of Ly$\alpha$ emitters (LAEs) have been noted between {\tt JWST} and ground-based facilities. We employ {\tt SPICE}, a suite of radiation-hydrodynamical simulations featuring different stellar feedback models, and investigate the impact of radiative transfer effects (e.g. extended emission, Ly$\alpha$-UV spatial offsets) and observational systematics (such as slit placement) on the measured Ly$\alpha$ EW$_0$. We perform radiative transfer of Ly$\alpha$ and UV photons for {\tt SPICE} galaxies to mimic slit spectroscopy for ground-based slits and {\tt JWST-MSA} pseudo-slits. We find that spatial Ly$\alpha$-UV offsets exist independently of feedback model, and are common ($>70$\% galaxies with $M_* > \rm 10^8 M_\odot$) with median values of $\approx0.07-0.11''$. The theoretical predictions from {\tt SPICE} are consistent with the observed spatial offset distribution. In addition, spatial Ly$\alpha$-UV offsets are identified as a major cause of loss of flux for {\tt JWST-MSA} type observations, with median pseudo-slit losses of $\approx65$\%, and $\approx30$\% cases suffering from $>95$\% pseudo-slit losses. Even in the absence of such spatial offsets, the presence of extended emission can cause median pseudo-slit losses of $40$\%, with $4$\% cases suffering from $>95$\% pseudo-slit losses. Finally, complex galaxy morphologies or misplaced {\tt JWST-MSA} pseudo-slit can lead to under-estimated UV continuum, resulting in spuriously high estimates of EW$_0$ from {\tt JWST} in $6-8$\% of galaxies. We compare the predictions from {\tt SPICE} to a sample of 25 galaxies with Ly$\alpha$ emission observations  from both the ground and from {\tt JWST}. The EW$_0^{\tt JWST}$ and the EW$_0^{\tt Ground}$ exhibit scatter in line with predictions from {\tt SPICE}, indicating that both physical and systematic effects are likely at play. 
\end{abstract}



\section{Introduction}
The Lyman alpha (Ly$\alpha$) emission line is a key tool in our quest to understand the high-redshift Universe. Ly$\alpha$ radiation is produced both via  recombinations around young stars and collisionally-excited hydrogen gas, and can be particularly bright \citep{1967partridge}. Due to its visibility and strength, the Ly$\alpha$ line has been used to spectroscopically confirm star-forming galaxies \citep[e.g.][]{Stark2010, Jung2020},
as a proxy for Lyman continuum (LyC) escape to pinpoint galaxies that drive cosmic reionisation \citep[e.g.][]{Verhamme2015, Marchi2017}, 
and to understand the physical properties of the high-redshift intergalactic medium \cite[IGM; for an overview see][]{Dijkstra2014, Ouchi2020}. 

The evolution of the fraction of Ly$\alpha$ emitters\footnote{A galaxy is classified as a LAE if its Ly$\alpha$ equivalent width, EW$_0$, is $>25$\AA, where EW$_0$ is evaluated as the ratio between the Ly$\alpha$ flux and the UV continuum of the galaxy.} (LAEs) relative to that of the UV continuum selected galaxies ($X_{\rm Ly\alpha}$) is often used to infer the IGM optical depth. To this aim, the distribution of intrinsic Ly$\alpha$ equivalent width in UV selected galaxies is assumed to be the one observed in the post-reionisation Universe, at $z \sim 5-6$ \citep[][hereafter N24]{Fontana2010,Stark2010,Pentericci2011,Jung2020,Napolitano2024}.
Indeed, several investigations \citep[e.g.][]{Pentericci2011, Caruana2014, Schenker2014} have reported a decline of $X_{\rm Ly\alpha}$ at high redshift, which becomes particularly rapid at $z>6$, suggesting an increasingly neutral universe. However, the interpretation of this trend remains debated due to potential biases and systematics in observing strategies. 

Conventional strategies to identify LAEs include narrow-band imaging \citep[e.g.][]{Ouchi2008,Zheng2016}, integral field unit \citep[IFU; e.g.][]{Wisotzki2018}, and slit spectroscopy \citep[e.g.][]{Pentericci2018}. As the vast majority of instruments employed for these searches are ground-based, the sensitivity of the observations is limited by sky background and atmospheric telluric lines, in particular at $z>7$. In addition, in the absence of Ly$\alpha$ the redshift confirmation of galaxies remains dubious, due to the lack of other features in the spectra of these galaxies. This renders the observed of the fraction of LAEs difficult to interpret and adds significant scatter to the available data. 
In the first two years of operation, the \texttt{James Webb} Space Telescope's Near InfraRed Spectrograph \citep[\texttt{JWST-NIRSpec},][]{Gardner2023, Jakobsen2022} has demonstrated its ability to successfully identify Ly$\alpha$ emission from high-redshift galaxies \citep[e.g.][]{Jung2023, Tang2023, Saxena2024}, enabling the study of the Ly$\alpha$ visibility evolution during the Epoch of Reionisation  \citep[EoR; e.g.][]{Chen2024, Jones2024, Nakane2024, Napolitano2024, Tang2024}. Most notably, {\tt JWST} allows for confirmation of redshifts of galaxies using optical bright emission lines (e.g. Balmer lines, [O III]). Therefore the absence of Ly$\alpha$ emission can be quantified in a very precise way along with the evaluation of $X_{Ly\alpha}$. 

Crucially, various studies have reported discrepancies between the Ly$\alpha$ rest-frame equivalent width measured by \texttt{JWST} and the estimates from ground-based telescopes \citep{Chen2024,Larson_2023,Tang2023}, including a case in which \texttt{JWST} has reported a non-detection, while strong Ly$\alpha$ emission has been measured by \texttt{VLT-MUSE} \citep{Jiang2023}. 
These discrepancies in turn translate into  $X_{\rm Ly\alpha}$ estimates from \texttt{JWST} significantly lower than those from ground-based observations at $z=5-6$. 
The loss of Ly$\alpha$ flux is mainly attributed to the small size ($0.12''\times0.46''$) of the micro-shutter assembly on \texttt{JWST}, which could potentially miss extended emission caused by resonant scattering of Ly$\alpha$ \citep{Verhamme2012,Smith2015,Byrohl2021,Smith2022,Yuxuan2024}. Another possible explanation is the presence of a spatial offset between the UV and Ly$\alpha$ emission, which would imply that if a pseudo-slit is oriented on the UV peak, the Ly$\alpha$ peak could fall outside of the collection area of the slit, leading to flux loss. In the latter respect, a few observational investigations have attempted to quantify the incidence and strengths of such spatial UV-Ly$\alpha$ offsets \citep{Hoag2019,Lemaux2021,Ning2024}, finding that offsets exist with medians of $\sim 0.1''$, and their strength tends to decrease at higher redshifts. Furthermore, \citet{Ning2024} report a potential positive correlation between the offset strength and the Ly$\alpha$ equivalent width, implying that stronger emitters could be more susceptible to flux losses. 

Interpreting the influx of ground- and space-based observations of galaxies during the EoR requires pinning down the magnitude and incidence of Ly$\alpha$-UV offsets and spatially-extended emission.
Attempts to forward-model slit-losses caused by spatial offsets \citep{Nakane2024} have shown that losses are expected to be $\sim20$\%. However, these estimates are based on isotropic 2D models of Ly$\alpha$ and UV emission, which most likely miss the complexity and strong directional dependencies associated to the multi-phase nature of the galactic interstellar medium (ISM) \citep[see also][]{Garel2021,Smith2022,Blaizot2023,Choustikov2024}. 
Here, we quantify the incidence and importance of Ly$\alpha$-UV offsets and spatially-extended emission in high redshift galaxies using {\tt SPICE} \cite[][hereafter B24]{bhagwat2024}, a suite of three cosmological radiation-hydrodynamic simulations with different stellar feedback prescriptions, in which the Ly$\alpha$ and UV properties of galaxies can be modeled while accounting for resonant scattering and radiative transfer through a complex multi-phase ISM. 

The paper is organised as follows: in Section~\ref{sec:Thoereticalmodelling} we describe the {\tt SPICE} simulations (\ref{sec:SPICE}), the Ly$\alpha$ radiation transfer computation (\ref{sec:Lya-RT}), and the construction of synthetic IFU observations (\ref{sec:IFUmocks}). In Section~\ref{sec:observations} we introduce observations of high redshift LAEs with ground-based facilities (\ref{sec:ground}) and \texttt{JWST} (\ref{sec:jwstobs}), and compare their Ly$\alpha$ EW$_0$ (\ref{sec:comparejwstvlt}). In Section~\ref{sec:simulationresult} we show theoretical predictions from {\tt SPICE}. Finally, in Section~\ref{sec:discussion} we discuss implications of our findings on interpretations of observations and summarise our key results.
\section{Theoretical modeling}
\label{sec:Thoereticalmodelling}
Here we briefly describe the simulations used for the analysis, as well as the methodology adopted to construct synthetic datacubes. 
\subsection{SPICE simulations}
\label{sec:SPICE}
We use {\tt SPICE} (B24), a suite of radiation-hydrodynamic simulations performed with the adaptive mesh refinement code {\tt RAMSES-RT} \citep{rosdahl2013,rosdahl2015}. 
The simulations are run in a cosmological box of length $10 {h^{-1}}$~cMpc, with
$512^{3}$ dark matter particles of mean mass $6.38 \times 10^{5} \mathrm{M_{\odot}}$. We adopt 
a  $\rm \Lambda CDM$ model with $\Omega_{\Lambda} = 0.6901$, $\Omega_{\rm b} = 0.0489$, $\Omega_{\rm m} = 0.3099$, $H_0 = 67.74$ \ \rm km~s$^{-1}$~Mpc$^{-1}$, $\sigma_8 = 0.8159$, and $n_{\rm s} = 0.9682$ \citep{Planck2016}. 

Metal line cooling is accounted for at $T>10^4 \rm K$ adopting {\tt CLOUDY} tables \citep{ferland1998cloudy}, while fine structure rates from \citet{rosen1995global} are employed for $T\leq10^4 \rm K$. 
The non-equilibrium ionisation states of H and He are advected while being fully coupled to the local ionising radiation \citep[see][]{rosdahl2013ramses}. 
{\tt SPICE} employs a star formation model with a variable star formation efficiency \citep{kretschmer2020forming} which depends on the local value of the gas turbulent Mach number and virial parameter, the latter being an indicator of the local stability. We refer the reader to B24 for more details of the model.
We adopt a Chabrier initial mass function \citep{chabrier2003galactic}, which results in a Supernova (SN) rate of 0.016 SN $\rm M^{-1}_\odot$.
SN mechanical feedback is implemented as in \cite{kimm2014escape} and  \cite{kimm2015towards}.
A unique feature of {\tt SPICE} is the use of three different supernova feedback models, resulting in three highly-contrasting star formation and feedback behaviours.
This is achieved by maintaining the implementation of the SN feedback, while varying the energy and timing of the SN explosions as follows (refer also to Table 2 of B24):
\begin{enumerate}
\item {\tt \textbf{bursty-sn}}: When a stellar particle becomes 10~Myr old, all SN explode in a single event injecting an energy of $2\times10^{51}$ ergs. 
\item {\tt \textbf{smooth-sn}}:  As (i), but SN events now happen between 3 and 40~Myr since the stellar particle birth.  
\item {\tt \textbf{hyper-sn}}: As (ii), but a fraction $f_{\rm HN}$ of SN explodes as hypernovae, with an energy of $10^{52}$ ergs. We adopt a metallicity dependent $f_{\rm HN} = {\rm max}[0.5 \times \ {\rm exp}(-Z_\ast/0.001), 0.01]$, with $Z_\ast$ stellar metallicity \citep{grimmett2020chemical}. The other SN events have an energy in the range $10^{50} - 2\times10^{51}$ ergs \citep{sukhbold2016core}. 
\end{enumerate}

We follow the on-the-fly radiative transfer of photons in five frequency bands, i.e. infrared (IR, $0.1-1 \ \rm eV$), optical ($1-13.6 \ \rm eV$) and three ionising UV bands (13.6-24.59~eV, 24.59-54.42~eV, and >54.42~eV). In addition to radiative feedback from photoionisation and photoheating, we also include radiation pressure from UV photons and radiation pressure on dust from IR and optical photons.

Spectral energy distributions taken from \texttt{BPASSv2.2.1} \citep{bpassv211,bpassv221} are employed to evaluate the stellar particles' luminosity based on their metallicity, age and mass. We adopt a dust number density $n_{\rm d} \equiv (Z/Z_\odot) n_{\rm HI}$, where $n_{\rm HI}$ is the neutral hydrogen number density. Photons interact with gas through multi-scattered radiation pressure and can be re-processed into the IR according to dust absorption and scattering opacities assigned to each cell (see \citealt{rosdahl2015scheme} and Table 3 in B24).

\subsection{Ly$\alpha$ radiative transfer}
\label{sec:Lya-RT}

We perform Ly$\alpha$ radiative transfer (Ly$\alpha$-RT) in post-process using the publicly available, resonant-line transfer code {\tt RASCAS} \citep{Michel_Dansac_2020}. This calculates the spatial and spectral diffusion of resonant-line photons using a Monte Carlo technique.

We model Ly$\alpha$ photon production from three emission channels: (i) recombination radiation from photo-ionised gas, (ii) Ly$\alpha$ cooling from collisionally-excited hydrogen, and (iii) direct injection from the stellar continuum. The emission from a recombination cascade from photo-ionised gas is modelled as in \citet{Cantalupo2008}, with the number of Ly$\alpha$ photons emitted per unit time in a gas cell of proper length $\Delta x$ given by
\begin{equation}
    \Dot{N}_{\rm Ly\alpha, rec} = n_{\rm e} n_{\rm p} \epsilon^{\rm B}_{\rm Ly\alpha}(T) \alpha^{\rm B}(T) (\Delta x)^3 ,
     \label{eq:recombination}
\end{equation}
where $n_{\rm e}$ and $n_{\rm p}$ are the electron and proton number densities, $T$ is the gas temperature, $\alpha^{\rm B}(T)$ is the case-B recombination coefficient \citep[Appendix A]{Hui1997}, and $ \epsilon^{\rm B}_{\rm Ly\alpha}(T)$ is the number of Ly$\alpha$ photons produced per recombination event. 
The rate of Ly$\alpha$ photons emitted by collisionally-excited gas cell is given by
\begin{equation}
    \Dot{N}_{\rm Ly\alpha, col} = n_{\rm e} n_{\rm HI} \left[ \frac{6.58\times10^{-18}}{T^{0.185}} \right] \left[ \frac{e^{-(4.86\times10^4)/T^{0.895}}}{h \nu_{0}} \right] (\Delta x)^3 ,
    \label{eq:collisions}
\end{equation}
where $\nu_{\rm 0} = 2.47\times10^{-15} s^{-1}$ is the rest frame frequency of a Ly$\alpha$ photon.  The rate of collisional excitations is taken from the best fit parameters of \citet[Supplementary  data Fig. S1]{2022KatzMgII}. The stellar continuum is modelled using the same spectral energy distributions of  {\tt SPICE}, and the Ly$\alpha$ photons emission rate of each star particle is estimated using a 2D interpolation in age and metallicity. 
 
The Ly$\alpha$ radiation from recombination and collisionally-excited gas is sampled with $N_{\rm ph,rec}=N_{\rm ph,col}=10^6$ photon packets, while for the stellar continuum we adopt $N_{\rm ph,stc} = 1.5\times10^7$ packets\footnote{We note that the larger number of photon packets employed to sample the stellar continuum is due to the larger number of stellar particles in comparison to gas cells, as well as to the wider spectral band that needs to be sampled. Details and convergence tests will be provided in Bhagwat et al. (in prep.).}.

\texttt{RASCAS} models the interaction of the emitted Ly$\alpha$ photons with the hydrogen, deuterium and dust contained within each gas cell, assuming a deuterium abundance of D/H~$=3\times10^{-5}$, while the dust number density is provided by {\tt SPICE} (see section~\ref{sec:SPICE} of B24 for details).
The Ly$\alpha$-RT calculation includes recoil due to deuterium, dust absorption (assuming a Small Magellanic Cloud composition\footnote{The choice of dust composition does not affect our results (\citealt{Costa2022}).}), as well as scattering with all three species.  
{\tt RASCAS} adopts the phase functions from \citet{Hamilton1940} and \citet{Dijkstra2008} for scattering of photons around the line centre, and Raleigh scattering in the line wings. 
Ly$\alpha$ photons are scattered by dust with a probability given by an albedo $a_{\rm dust} = 0.32$ \citep{Li2001} following a Henyey-Greenstein phase function \citep{Henyey1941} with an asymmetry parameter $g=0.73$.  
To reduce computational overhead in regions of high optical depth ($\gg 10^3$), {\tt RASCAS} implements a core-skipping algorithm (see \citealt{Smith2015}) which shifts the photons to the line wings, facilitating their escape. 
Finally, {\tt RASCAS} employs the "peeling off" algorithm \citep{Zheng2002,Whitney2011} to collect Ly$\alpha$ flux along a given line-of-sight.  We post-process all {\tt SPICE} galaxies with $M_* > \rm 10^8 M_\odot$ at $z=5,6,7$ which have UV magnitudes of $-23 < M_{\rm UV} < -16$ with median $M_{\rm UV}=-18.3$ and $\beta_{\rm UV}$ ranging between $-3 < \beta_{\rm UV} < -1.5$.

\subsection{Synthetic IFU datacubes and associated observables}
\label{sec:IFUmocks}
To compare the Ly$\alpha$ properties of {\tt SPICE} galaxies to observations, we construct synthetic IFU datacubes of side length $3R_{\rm vir}$, where $R_{\rm vir}$ is the virial radius, centered on the halo. We follow the Ly$\alpha$-RT of photons produced from the three emission channels mentioned in Section~\ref{sec:Lya-RT}.
We adopt $N\times N$ spatial bins of resolution $\Delta \theta$, and $N_{\lambda}$ spectral bins of resolution $\Delta \lambda$ along the line-of-sight. Each 3D IFU datacube is thus constructed to have two spatial axes with $\Delta x = 0.05''$, and one wavelength axis with a rest-frame $\Delta \lambda \approx 0.13$\AA.

Each photon packet contributes a luminosity of $L_{{\rm Ly\alpha},i}/N_{{\rm ph},i}$, where $L_{{\rm Ly\alpha},i}$ is the total Ly$\alpha$ luminosity produced via channel $i$. The probability that a photon escapes to an observer positioned at a luminosity distance $D_{\rm L}$ within a given wavelength bin is $P(\mu) e^{-\tau_{\rm esc} (\lambda)}$, where $\tau_{\rm esc} (\lambda)$ is the optical depth between the scattering event and the edge of the computational domain. The total flux within each pixel in the datacube is written as
\begin{equation}
    F_{\rm Ly\alpha, pixel} = \frac{L_\lambda/N_{\rm ph}}{4 \pi D_{\rm L}^2 (1+z)} \sum P(\mu) e^{-\tau_{\rm esc}(\lambda)},
\end{equation}
where $L_\lambda = L_{\rm Ly\alpha} / [\Delta \lambda (1+z)^{-1}]$, such that the flux in each pixel is integrated over all scattering events from all photon packets. 

Calculations for collisional and recombination Ly$\alpha$ are carried out in the range $1205-1225$\AA\ and the stellar continuum is calculate in the range $800-2600$\AA. Henceforth, we refer to the UV continuum as flux estimated in range $1400-2600$\AA. For each halo, we calculate the datacubes along 12 sightlines, where 4 are oriented at an angle of [0, 30, 60, 90] degrees to the halo's angular momentum vector (fixed $\phi$ with random $\theta$), while the rest are drawn with random $\phi$ and $\theta$.  

From the IFU datacubes we evaluate the equivalent width (EW) and flux of Ly$\alpha$ and UV radiation. 
The flux is collected in apertures with sizes $0.2''\times0.46''$  and $0.7''\times8''$  to resemble the slit dimension on \texttt{JWST-NIRSpec} and ground-based settings, respectively. Note that ground based spectroscopy was performed using variable slit sizes, with width  from $0.7''$ to $1.0''$ and length which was in some cases also greater than 10$''$ \citep{Pentericci2018,Schenker2014,Stark2011}. While we always center the ground-based slits at the peak of the UV continuum, the  {\tt JWST-MSA} pseudo-slits are placed in two ways: either centered at the peak of the UV, or  centered with offsets between $0''$ and $0.15''$ with respect to the UV peak in random directions to account for positional uncertainty in the MSA placement.  This choice is motivated by the real offset in the observational data (see next section).
We fit a power law to the UV continuum to obtain the $\beta$ slope of a galaxy, and use the fit is used to estimate the continuum emission at the Ly$\alpha$ line center ($F^{\rm cont}_{\rm Ly\alpha}$). The rest-frame equivalent width (EW$\rm_0$) for Ly$\alpha$ is then measured as:
\begin{equation}
    {\rm EW_0} = \frac{F_{\rm Ly\alpha}}{F^{\rm cont}_{\rm Ly\alpha} (1 + z)}.
\end{equation}
Finally, we quantify the spatial offsets, $d_{\rm Ly\alpha-UV}$, between Ly$\alpha$ and UV by calculating the distance between the peak pixels in the respective surface brightness (SB) maps. We note that while this study focuses on the physical scenarios and systematics affecting interpretation of Ly$\alpha$ emission, a detailed study of Ly$\alpha$ properties of {\tt SPICE} galaxies including radial surface brightness profiles, luminosities, line profiles will follow in a companion work (Bhagwat et al. in prep.).
\section{Observational data}
\label{sec:observations}
As we aim to investigate potential systematics due to slit loss effects when comparing Ly$\alpha$ rest-frame equivalent widths from \texttt{JWST-NIRSpec} and ground-based measurements, we compile the largest possible sample of galaxies with Ly$\alpha$ measurements both from ground and space (i.e. \texttt{JWST-NIRSpec}) instrumentation. To minimise any possible  calibration issue, we compare the Ly$\alpha$ EW$_0$ rather than the emission line flux. 

\subsection{Ground-based observations}
\label{sec:ground}
We  assemble a catalog of galaxies which were observed with the aim of detecting Ly$\alpha$ emission from ground-based facilities  (\texttt{MUSE} and \texttt{FORS2} on the \texttt{VLT}, and \texttt{DEIMOS} and \texttt{MOSFIRE} on \texttt{Keck})  in the past $\sim 15$ years, targeting  the GOODS-South, GOODS-North, EGS, UDS, COSMOS, and Abell 2744 fields. We consider the published catalogs with their spectroscopic redshift $z_{\rm spec}$, Ly$\alpha$ EW$_0$ measurements, or upper limits, as presented in \cite{Pentericci2018}, \cite{Jung2020}, \cite{Richard2021}, \cite{Schmidt2021}, \cite{Jung2022}, \cite{Bacon2023}, and \cite{Napolitano2023}.
We ensure that there are no systematics related to the ground-based instruments used and consider sources at $z_{\rm spec}$ $\geq$ 5, to match the redshifts probed by the {\tt SPICE}.
\begin{figure*}
    \centering
    \includegraphics[width=\textwidth]{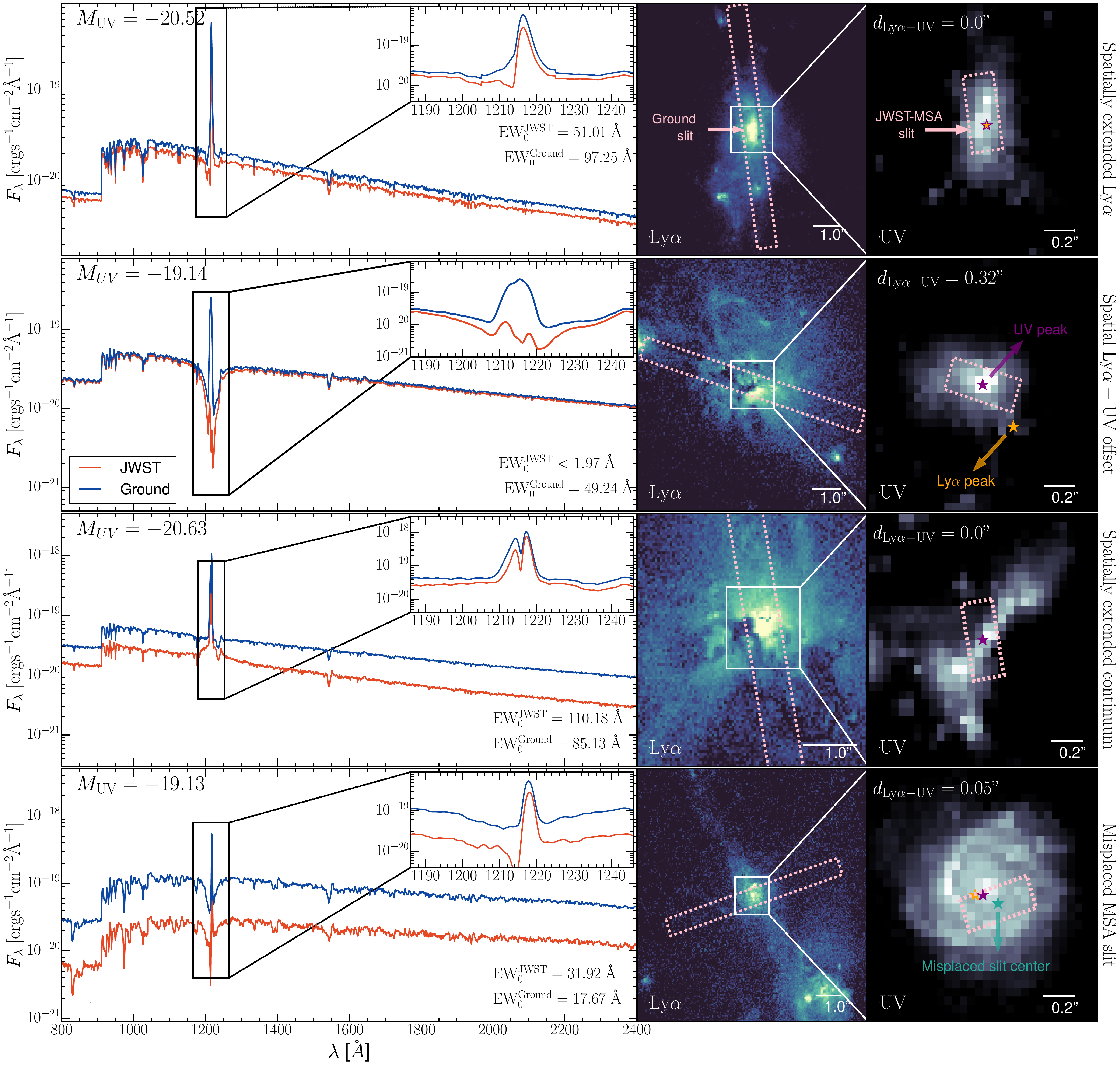}
    \caption{Four scenarios for EW$_0$ mismatch: {\it Left column:} Spectra extracted with ground-based slit (blue) and \texttt{JWST-MSA} pseudo-slit (red) from a reference galaxy in {\tt SPICE}; the inset shows the spectra in the range 1180-1250~\AA\ rest-frame. $M_{\rm UV}$ listed in each panel is the UV magnitude calculated between 1400-1500~\AA, while EW$_0$ is the Ly$\alpha$ equivalent width (see section~\ref{sec:IFUmocks}).  
    {\it Middle column:} Ly$\alpha$ emission from the galaxy with the ground-based slit overplotted (dotted line).
    {\it Right column:} zoomed-in region on the UV continuum of the galaxy with the \texttt{JWST-MSA} pseudo-slit overplotted (dotted line). The position of the peak of the emission in UV (purple star) and Ly$\alpha$ (orange star) is shown to highlight the spatial Ly$\alpha$-UV offset ($d_{\rm Ly\alpha-UV}$) .
    The rows refer to various cases in which a mismatch arises between the EW$_0$ measured by {\tt JWST} and by ground-based telescopes; from top to bottom:
    extended Ly$\alpha$ emission in which the peaks of the Ly$\alpha$ and UV emission coincide, a large spatial Ly$\alpha$-UV offset of $d_{\rm Ly\alpha-UV}=0.32''$, \texttt{JWST} missing UV continuum due to complex galaxy morphologies with a spatially extended continuum,
    and \texttt{JWST} missing UV continuum due to errors in the pseudo-slit placement, with the true pseudo-slit position shown as green star.}
    \label{fig:OffsetsSBmaps}
\end{figure*}
\subsection{\texttt{JWST} observations}
\label{sec:jwstobs}
We search for \texttt{JWST-NIRSpec} PRISM observations of targets which matched the sky position of the galaxies from the ground based catalog. To allow for positional uncertainties between different catalogs, we match objects using a 0.5'' radius threshold. We employ all the spectra available from the CEERS data release \citep{ArrabalHaro2023}, from the JADES-DR3 \citep{DEugenio2024} and from the DAWN \texttt{JWST} Archive\footnote{https://dawn-cph.github.io/dja/} \citep[DJA-Spec,][]{Heintz2024}. In particular, the observations include  the Abell 2744 \citep[PID 2561, 2756,][and Mascia et al., submitted]{Bezanson2022}, JADES GOODS-North \citep[GTO 1181, 1211,][]{Bunker2023}, JADES GOODS-South \citep[GTO 1180, 1210, PID 3215, 6541,][]{Bunker2023}, EGS \citep[ERS 1345,][]{ArrabalHaro2023}, COSMOS and UDS \citep[PID 2565,][]{Nanayakkara2024} fields.
We visually inspect all the spectra and the $z_{\rm spec}$ provided by JADES-DR3 and DJA-Spec, while for CEERS we use the catalog presented in \cite{Napolitano2024}. We also required the difference between the ground based and \texttt{JWST} spectroscopic redshift solutions to be less than 0.05 of each other (considering the low resolution of the PRISM observations). In total we obtained 60 matches.

For each \texttt{JWST} spectrum, we measure the UV magnitude, $M_{\rm UV}$, in the 1400--1500\AA\ rest-frame range without correcting for dust. The selected galaxies have a UV magnitude in the range $-21< M_{\rm UV} < -18$. The Ly$\alpha$ EW$_0$ and uncertainties are derived as detailed in \cite{Napolitano2024}. When the Ly$\alpha$ is not detected, we use the typically adopted 3EW$_{\rm 0,lim}$ as an upper limit, where EW$_{\rm 0,lim}$ is the lowest value that can be measured according to Equation (1) of \cite{Napolitano2024}, which takes into account the spectral resolution and the noise spectrum. 

\subsection{\texttt{JWST} and ground-based equivalent widths}
\label{sec:comparejwstvlt}
Out of the 60 galaxies selected above, the Ly$\alpha$ EW$_0$  from the \texttt{JWST} and ground based observations can be  compared in a meaningful way in 25 cases, which include: (i) All galaxies with EW$_0$ measurements from both  instruments (14 sources). (ii) All galaxies with a EW$_0$ measurement  from \texttt{JWST} and for which the 3$\sigma$ upper limit from ground-based instruments is below the \texttt{JWST} measure. In these cases we can conclude that the \texttt{JWST} measure is above the ground-based one (6 sources). (iii) All galaxies with a ground-based EW$_0$ measurement and for which Ly$\alpha$ is not detected in the \texttt{JWST} spectrum and its 3$\sigma$ upper limit is below the ground-based EW$_0$. In these cases we can conclude that the ground-based measure is above the \texttt{JWST} one (5 sources). 
In all other cases, for example when both the \texttt{JWST} and ground-based spectra only give upper limits, the comparison is inconclusive.

Finally utilizing the "slitlet viewer" tool provided in the DJA archive, we check for the positional offset between the center of the MSA pseudo-slit and the peak of the UV continuum emission for the 25 sources just discussed.  We found  positional offsets varying  between 0$''$ and 0.15$''$.\\
 
\section{Results}
\label{sec:simulationresult}
\texttt{SPICE} galaxies often exhibit a mismatch between the EW$_0$ that would be measured by {\tt JWST} and ground-based telescopes. We identified four main origin scenarios for such a mismatch: (i) spatially extended Ly$\alpha$ emission, (ii) spatial offsets between Ly$\alpha$ emission and the UV continuum, (iii) spatially extended young stellar populations and (iv) MSA misplacements.

In Figure~\ref{fig:OffsetsSBmaps} we illustrate these scenarios, using the ground-based slit (see Section~\ref{sec:IFUmocks}) as a reference for the "true" measured EW$_0$. 
The system in the top row shows a spatially extended Ly$\alpha$ nebula in which the peak of the Ly$\alpha$ and UV emission coincide, i.e. $d_{\rm Ly\alpha-UV}=0''$. While the {\tt JWST-MSA} pseudo-slit is able to capture the UV continuum to the same extent as the ground-based slit, the former misses part of the Ly$\alpha$ flux, leading to an underestimation of the EW$_0$ by a factor of $\approx2$.
The second row illustrates a case in which there is a significant spatial Ly$\alpha$-UV offset, with $d_{\rm Ly\alpha-UV}=0.32''$. While the UV continuum is captured by both slit and MSA pseudo-slit (as in the previous case), the majority of the Ly$\alpha$ emission lies outside of the {\tt JWST-MSA} pseudo-slit boundaries due to the spatial offset. This effect translates into the estimation of a large Ly$\alpha$ EW$_0$ with ground-based facilities and a non-detection of Ly$\alpha$ with {\tt JWST} (similar to the case reported by \citealt{Jiang2023}).
The third row illustrates the case in which the continuum is produced by a system with spatially extended star formation, but without the presence of a spatial Ly$\alpha$-UV offset. While the {\tt JWST-MSA} pseudo-slit captures the majority of the Ly$\alpha$ emission, it fails to capture the full extent of the continuum emission. This leads to an estimate of EW$_0$ with {\tt JWST} which is higher than the ground-based one, as the slit captures the full continuum emission. 
In this final scenario, if the pseudo-slit barycenter is not correctly placed at the peak of the UV continuum (bottom row), the small size of the {\tt JWST-MSA} pseudo-slit can miss a large portion of the UV continuum flux, resulting in a lower continuum estimate at the Ly$\alpha$ wavelength. As a consequence, \texttt{JWST} would overestimate EW$_0$ even if the Ly$\alpha$ emission were perfectly captured, while ground-based observations would instead correctly capture the full UV continuum as in the previous case, consequently, \texttt{JWST} would overestimate EW$_0$ as compared to ground-based slit. Therefore, both the systematics of observations and the underlying physical scenarios can affect the interpretation of measured EW$_0$. 

\begin{figure}
    \includegraphics[width=0.47\textwidth]{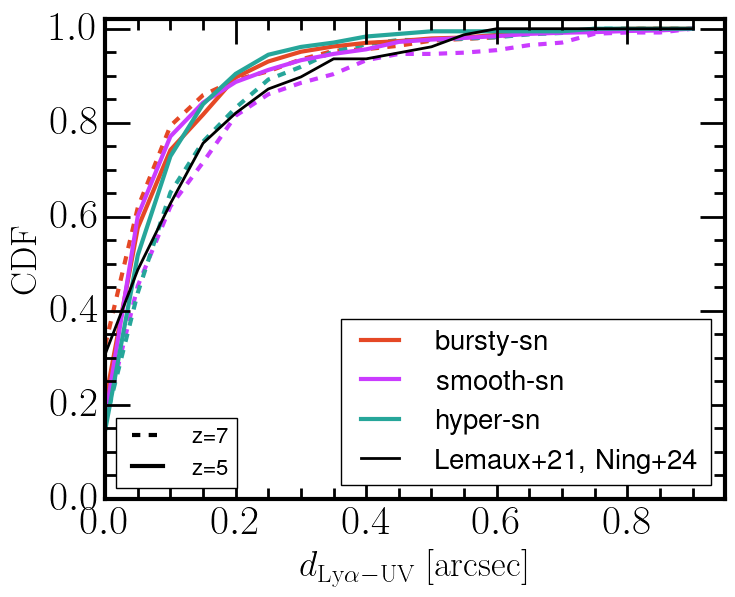}
    \caption{Cumulative distribution functions (CDF) of spatial Ly$\alpha$-UV offsets, $d_{\rm Ly\alpha-UV}$, for the three feedback models at $z=7$ (dashed lines) and 5 (solid).  The black solid line shows the cumulative distribution function of offsets obtained using the datasets of \citet{Lemaux2021} and \citet{Ning2024}. Predictions from {\tt SPICE} are consistent with observed spatial offsets independent of feedback model. 
    }
    \label{fig:OffsetCDF}
\end{figure}

As the presence of offsets can potentially lead to JWST missing a large portion of  Ly$\alpha$ flux leading to a non-detection of Ly$\alpha$ emission, it is key to investigate their incidence and strength. In Figure~\ref{fig:OffsetCDF} we show the cumulative distribution function of $d_{\rm Ly\alpha-UV}$ at $z$=5 and 7\footnote{We find similar results at $z$=6, but they are not shown to avoid overcrowding.}. For all models and at all redshifts analysed, over the full sample, we find non-zero offsets in $>70$\% galaxies, with median values of $0.07\pm0.06'' (0.07\pm0.05''),0.08\pm0.06'' (0.10\pm0.07'')$ and $0.077\pm0.08''  (0.11\pm0.08'')$  for {\tt bursty-sn}, {\tt smooth-sn} and {\tt hyper-sn}, respectively at $z=5$ (7). For the sub-sample of galaxies with non-zero offsets, the median values are $0.10\pm0.05'' (0.11\pm0.05''),0.08\pm0.05'' (0.12\pm0.05'')$ and $0.11\pm0.07'' (0.12\pm0.07'')$  for {\tt bursty-sn}, {\tt smooth-sn} and {\tt hyper-sn}, respectively at $z=5$ (7). We note a slight redshift evolution in the {\tt smooth-sn} and {\tt hyper-sn} models, where the median offset decreases between $z=7$ and 5, while {\tt bursty-sn} shows no evolution. Median results are consistent with previous observations of high redshift galaxies \citep{Hoag2019,Ribeiro2020,Khusanova2020,Lemaux2021,Ning2024,2024Navarre}. However, some of the observations  show a mild decrease in offsets with increasing redshift, in contrast to what we obtain using the {\tt smooth-sn} and {\tt hyper-sn} models.

\begin{figure}          
    \includegraphics[width=0.47\textwidth]{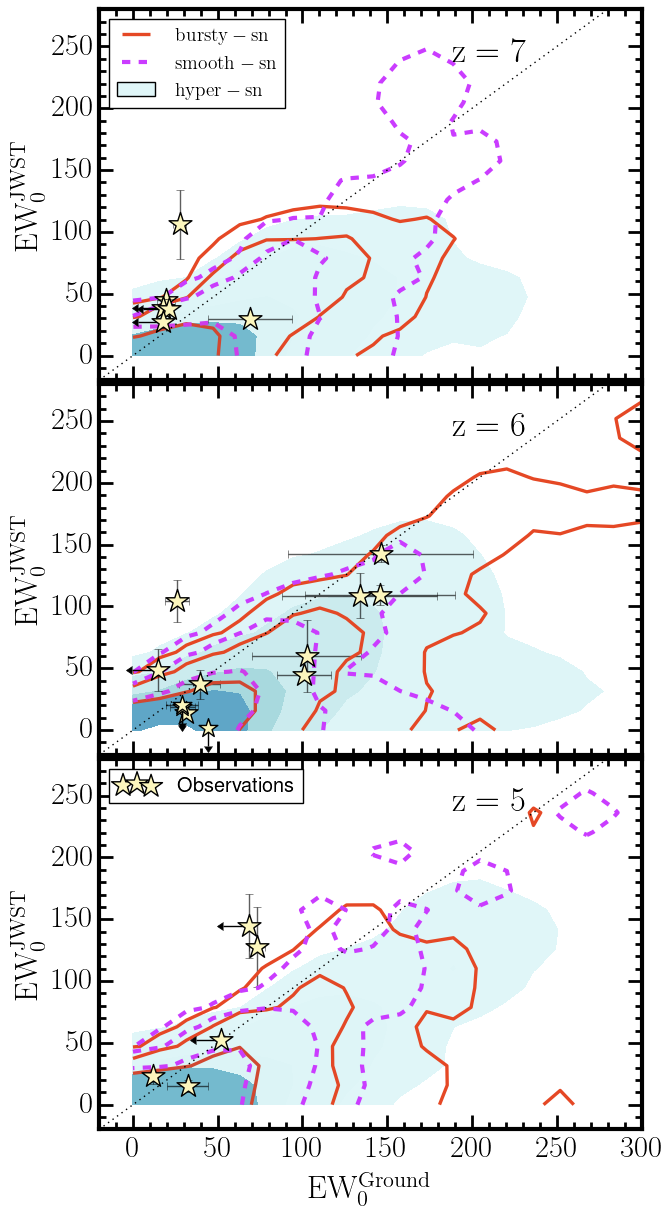}
    \caption{Ly$\alpha$ EW$_0$ measured by {\tt JWST} as a function of the one  measured by ground-based facilities for the three feedback models at $z=7$ (top panel), $6$ (middle), and $5$ (bottom). Contours show regions covering 68\%, 95\% and 99\% of all points. The black dotted line is the 1:1 relation, while the star symbols represent the 25 sources for which the comparison between EW$_0^{\tt JWST}$ and EW$_0^{{\tt Ground}}$ is conclusive (see Section~\ref{sec:observations} for details). Arrows refer to 3$\sigma$ upper limits for non-detections. The majority of values derived from {\tt SPICE} lie below the 1:1 line irrespective of the feedback model, implying a deficit of EW$_0$ as measured by {\tt JWST}.
    }
    \label{fig:EWvsEW}
\end{figure}

To assess the effect of slit-sizes on the estimated EW$_0$ of Ly$\alpha$, in Figure~\ref{fig:EWvsEW} we show the EW$_0$ measured using the {\tt JWST-MSA} pseudo-slit as a function of the value measured with a ground-based slit, for pseudo-slits centered on the peak of the UV. 
At $z=7$ (top panel), in all models, the majority of galaxies has EW$_0^{\tt JWST} \ll $EW$_0^{\tt Ground}$ ($\approx98$\% in total), while only a small fraction has EW$_0^{\tt JWST}>$ EW$_0^{\tt Ground}$ ($\approx2$\%). At $z=6$ (middle panel), the difference in the estimates reduces slightly, with all models moving closer to the 1:1 line (equal EW$_0$), and the fraction of galaxies with EW$_0^{\tt JWST}>$EW$_0^{\tt Ground}$ reducing to $1\%$. By $z=5$ (bottom panel) this fraction has become negligible, while most galaxies still lie below the 1:1 line. The overall behaviour is maintained also with the inclusion of pseudo-slits misalignments, but the fraction of galaxies with EW$_0^{\tt JWST}>$ EW$_0^{\tt Ground}$ increases to 9\% (8\%) at $z=7 (5)$. In the case where we include positional offsets in pseudo-slit placement, the scatter in the contours increases for all models,  such that $\approx6$\% galaxies cross above the 1:1 at all redshifts. 

In Figure~\ref{fig:EWvsEW} we also report the 25 galaxies selected in Section~\ref{sec:observations} at their respective redshifts. We find 14 cases where galaxies lie below the 1:1 line, with {\tt JWST} measurements within a factor of $1-2$ of ground-based measurements; 6 galaxies have EW$_0^{\tt JWST}>$ EW$_0^{\tt Ground}$ which could potentially be due to pseudo-slit position not being centered on the peak of the UV continuum; and in 5 cases {\tt JWST-MSA} gives only a 3-sigma upper limit with confirmed strong Ly$\alpha$ detections from the ground. The observed data exhibits a scatter as predicted by {\tt SPICE} ,therefore, all the physical and systematic effects illustrated in Figure~\ref{fig:OffsetsSBmaps} could potentially be at play. We defer a more detailed investigation of the relative importance of the various mechanisms to a future study once more observational data are available.

\section{Summary and Conclusions}
\label{sec:discussion}
In this paper we use the SPICE simulations to show that the Ly$\alpha$ visibility can be strongly influenced by Ly$\alpha$ radiative transfer effects, and highlight the importance of considering spatial Ly$\alpha$-UV offsets when performing MSA spectroscopic observations with {\tt JWST}. Systematics such as pseudo-slit misalignments are also important to consider while interpreting observations. The combined information of rest-frame UV photometry and slit spectra can potentially correct for systematics.

We find that extended Ly$\alpha$ emission is present around every simulated galaxy in our sample \citep[also shown in][]{Verhamme2012,Byrohl2021,Smith2022,Yuxuan2024}. If there is no spatial Ly$\alpha$-UV offset, the emission radial profile determines whether EW$_0$ is correctly captured by {\tt JWST} or not, since it will be underestimated if the Ly$\alpha$ emission extends beyond the pseudo-slit size. Meanwhile, the larger slit sizes on the ground-based facilities allow to capture Ly$\alpha$ emission out to larger spatial scales.
In the {\tt bursty-sn} model, for galaxies without a spatial Ly$\alpha$-UV offset, pseudo-slit losses are estimated to be $\approx40$\% at all redshifts with $\approx$4\% ($7.2$\%) of Ly$\alpha$ emitting galaxies suffering pseudo-slit losses of $>95$\% when observed by {\tt JWST} at $z=7$ ($z=5$). For  the {\tt smooth-sn} and {\tt hyper-sn} models, the pseudo-slit losses mildly evolve between $40$\% at $z=7$ to $35$\% at $z=5$, with $\approx$4\% galaxies suffering pseudo-slit losses of $>95$\%.

The Ly$\alpha$ EW$_0$, however, are mostly affected by spatial Ly$\alpha$-UV offsets. For LAEs with spatial offsets, the pseudo-slit losses are estimated to be $\approx65$\% at all redshifts. This result appears to be robust as it holds throughout all \texttt{SPICE} simulations, i.e. irrespective of SN feedback model. Out of these galaxies, the fraction which suffer $>95$\% pseudo-slit loss in the {\tt bursty-sn} models increases from 30\% at $z=7$ to 38\% at $z=5$, while it is $31$\% ($30$\%) and $43$\% ($44$\%) at $z$ = 5 (7) for {\tt smooth-sn} and {\tt hyper-sn} models, respectively. Finally, $\approx22-24$\% of all LAEs show consistent EW$_0$ from both {\tt JWST-MSA} and ground-based slits. 

These results are obtained assuming that the {\tt JWST-MSA} pseudo-slit is perfectly centered on the peak of the UV. This is not always the case, since  the positions could be slightly misaligned to better optimize the total number of observed targets in the {\tt JWST-MSA} configuration. If we include such placement uncertainties, of the order of 0.0$''$--0.15$''$, as derived from the observations considered (see section~\ref{sec:observations}), the fraction of galaxies with such high pseudo-slit losses drops by $5-7$\% at all redshifts. These galaxies end up with overestimated EW$_0$ values (as shown in the bottom row of Figure~\ref{fig:OffsetsSBmaps}). Overall, the EW$_0$ estimated by {\tt JWST-MSA} increases relative to the ground-based values by a median factor of $\approx 4-5$ in the presence of slightly misaligned pseudo-slit positions for $\approx25$\% of galaxies. This effect is independent of feedback model and redshift considered. 

Due to both pseudo-slit losses and  misalignments, the scatter in the measured EW$_0$ distribution will lead to an uncertainty in the statistical visibility of Ly$\alpha$ that must be considered when using it to infer the neutral fraction of the IGM. Intrinsically, about $\approx30$\% LAEs will be missed by {\tt JWST-MSA} as predicted by {\tt SPICE} with the remainder being measured with mild to moderate pseudo-slit losses.

We summarize our key conclusions as follows: 
\begin{itemize}
    \item Radiative transfer effects leading to e.g. extended emission and Ly$\alpha$-UV offsets, as well as observational systematics (e.g. off-centered pseudo-slit positions) affect EW$_0$ measurements from {\tt JWST} introducing scatter in the EW$_0$ distributions.
    \item We find that $>70$\% {\tt SPICE} galaxies have non-zero spatial Ly$\alpha$-UV offsets, with a median offset of $0.07''$ ($0.07''$), $0.08''$ ($0.10''$) and $0.08''$ ($0.11''$) for  the {\tt bursty-sn}, {\tt smooth-sn} and {\tt hyper-sn} models, respectively, at $z$ = 5 ($z$ = 7). 
    \item The distribution of offsets shows only a mild redshift evolution for the {\tt smooth-sn} and {\tt hyper-sn} models, where the median offset decreases between $z$ = 7 and $z$ = 5, while {\tt bursty-sn} exhibits no redshift evolution in agreement with observations.
    \item In cases without spatial Ly$\alpha$-UV offsets, extended Ly$\alpha$ emission can cause median pseudo-slit losses of $\approx40$\% with $\approx4-7$\% of LAEs suffering from $>95$\% losses.
    \item Spatial Ly$\alpha$-UV offsets are the main cause of Ly$\alpha$ flux being underestimated from {\tt JWST} observations, with median pseudo-slit losses of $\approx65$\% in galaxies with significant spatial offsets. About $30-40$\% cases suffer from $>95$\% losses depending on redshift and feedback model.
    \item Complex galaxy morphologies as in the case of mergers or extended star-forming regions along with misplaced placement of the {\tt JWST-MSA} pseudo-slits can lead to the UV continuum being under-sampled. In this scenario, the EW$_0$ estimated from {\tt JWST} will exceed those estimated from the ground. About 2\% ($<1$\%) galaxies exhibit EW$_0^{\tt JWST} > {\rm EW}_0^{\tt Ground}$ without MSA placement offsets at $z$ = 7 ($z$ = 5) with the numbers increasing to  $\approx6\%$ (8\%) galaxies with the inclusion of MSA placement offsets.   

While pseudo-slit losses add scatter to Ly$\alpha$ EW$_0$, the physical phenomena that lead to such losses also are sensitive to the state of the ISM and the CGM of high-z LAEs. 
Future observations such as those from the upcoming JWST large program CAPERS  \citep[The CANDELS-Area Prism Epoch of Reionization Survey;][]{Dickinson2024} and RUBIES will provide larger samples of high redshift galaxies BOASTING  both JWST and ground-based observations of the Ly$\alpha$ emission line. Such samples  will be key to constrain the incidence and origin of extended emission and spatial offsets that affect the interpretation of Ly$\alpha$ emission visibility during cosmic reionisation.
\end{itemize}
\section*{Acknowledgements}
L.N. and L.P. acknowledge support from the PRIN 2022 MUR project 2022CB3PJ3 - First Light And Galaxy aSsembly (FLAGS) funded by the European Union – Next Generation EU. A.B would like to thank Eileen Herwig and Christian Partmann for useful discussions, Jeremy Blaizot and Leo-Michel Dansac for developing and maintaining {\tt RASCAS}. This work made use of publicly available software packages:  {\tt matplotlib} \citep{hunter2007matplotlib}, {\tt numpy} \citep{van2011numpy}, {\tt scipy} and \citep{scipy}. A.B extends a thank you to the community of developers and those maintaining all of these packages.
\section*{Data Availability}


\bibliographystyle{mnras}
\bibliography{biblio} 

\begin{thebibliography}{}
\makeatletter
\relax
\def\mn@urlcharsother{\let\do\@makeother \do\$\do\&\do\#\do\^\do\_\do\%\do\~}
\def\mn@doi{\begingroup\mn@urlcharsother \@ifnextchar [ {\mn@doi@}
  {\mn@doi@[]}}
\def\mn@doi@[#1]#2{\def\@tempa{#1}\ifx\@tempa\@empty \href
  {http://dx.doi.org/#2} {doi:#2}\else \href {http://dx.doi.org/#2} {#1}\fi
  \endgroup}
\def\mn@eprint#1#2{\mn@eprint@#1:#2::\@nil}
\def\mn@eprint@arXiv#1{\href {http://arxiv.org/abs/#1} {{\tt arXiv:#1}}}
\def\mn@eprint@dblp#1{\href {http://dblp.uni-trier.de/rec/bibtex/#1.xml}
  {dblp:#1}}
\def\mn@eprint@#1:#2:#3:#4\@nil{\def\@tempa {#1}\def\@tempb {#2}\def\@tempc
  {#3}\ifx \@tempc \@empty \let \@tempc \@tempb \let \@tempb \@tempa \fi \ifx
  \@tempb \@empty \def\@tempb {arXiv}\fi \@ifundefined
  {mn@eprint@\@tempb}{\@tempb:\@tempc}{\expandafter \expandafter \csname
  mn@eprint@\@tempb\endcsname \expandafter{\@tempc}}}

\bibitem[\protect\citeauthoryear{{Arrabal Haro} et~al.,}{{Arrabal Haro}
  et~al.}{2023}]{ArrabalHaro2023}
{Arrabal Haro} P.,  et~al., 2023, \mn@doi [\apjl] {10.3847/2041-8213/acdd54},
  \href {https://ui.adsabs.harvard.edu/abs/2023ApJ...951L..22A} {951, L22}

\bibitem[\protect\citeauthoryear{{Bacon} et~al.,}{{Bacon}
  et~al.}{2023}]{Bacon2023}
{Bacon} R.,  et~al., 2023, \mn@doi [\aap] {10.1051/0004-6361/202244187}, \href
  {https://ui.adsabs.harvard.edu/abs/2023A&A...670A...4B} {670, A4}

\bibitem[\protect\citeauthoryear{{Bezanson} et~al.,}{{Bezanson}
  et~al.}{2022}]{Bezanson2022}
{Bezanson} R.,  et~al., 2022, \mn@doi [arXiv e-prints]
  {10.48550/arXiv.2212.04026}, \href
  {https://ui.adsabs.harvard.edu/abs/2022arXiv221204026B} {p. arXiv:2212.04026}

\bibitem[\protect\citeauthoryear{{Bhagwat}, {Costa}, {Ciardi}, {Pakmor}  \&
  {Garaldi}}{{Bhagwat} et~al.}{2024}]{bhagwat2024}
{Bhagwat} A.,  {Costa} T.,  {Ciardi} B.,  {Pakmor} R.,   {Garaldi} E.,  2024,
  \mn@doi [\mnras] {10.1093/mnras/stae1125}, \href
  {https://ui.adsabs.harvard.edu/abs/2024MNRAS.531.3406B} {531, 3406}

\bibitem[\protect\citeauthoryear{{Blaizot} et~al.,}{{Blaizot}
  et~al.}{2023}]{Blaizot2023}
{Blaizot} J.,  et~al., 2023, \mn@doi [\mnras] {10.1093/mnras/stad1523}, \href
  {https://ui.adsabs.harvard.edu/abs/2023MNRAS.523.3749B} {523, 3749}

\bibitem[\protect\citeauthoryear{{Bunker} et~al.,}{{Bunker}
  et~al.}{2023}]{Bunker2023}
{Bunker} A.~J.,  et~al., 2023, \mn@doi [arXiv e-prints]
  {10.48550/arXiv.2306.02467}, \href
  {https://ui.adsabs.harvard.edu/abs/2023arXiv230602467B} {p. arXiv:2306.02467}

\bibitem[\protect\citeauthoryear{{Byrohl} et~al.,}{{Byrohl}
  et~al.}{2021}]{Byrohl2021}
{Byrohl} C.,  et~al., 2021, \mn@doi [\mnras] {10.1093/mnras/stab1958}, \href
  {https://ui.adsabs.harvard.edu/abs/2021MNRAS.506.5129B} {506, 5129}

\bibitem[\protect\citeauthoryear{{Cantalupo}, {Porciani}  \&
  {Lilly}}{{Cantalupo} et~al.}{2008}]{Cantalupo2008}
{Cantalupo} S.,  {Porciani} C.,   {Lilly} S.~J.,  2008, \mn@doi [\apj]
  {10.1086/523298}, \href
  {https://ui.adsabs.harvard.edu/abs/2008ApJ...672...48C} {672, 48}

\bibitem[\protect\citeauthoryear{{Caruana}, {Bunker}, {Wilkins}, {Stanway},
  {Lorenzoni}, {Jarvis}  \& {Ebert}}{{Caruana} et~al.}{2014}]{Caruana2014}
{Caruana} J.,  {Bunker} A.~J.,  {Wilkins} S.~M.,  {Stanway} E.~R.,  {Lorenzoni}
  S.,  {Jarvis} M.~J.,   {Ebert} H.,  2014, \mn@doi [\mnras]
  {10.1093/mnras/stu1341}, \href
  {https://ui.adsabs.harvard.edu/abs/2014MNRAS.443.2831C} {443, 2831}

\bibitem[\protect\citeauthoryear{Chabrier}{Chabrier}{2003}]{chabrier2003galactic}
Chabrier G.,  2003, Publications of the Astronomical Society of the Pacific,
  115, 763

\bibitem[\protect\citeauthoryear{{Chen}, {Stark}, {Mason}, {Topping},
  {Whitler}, {Tang}, {Endsley}  \& {Charlot}}{{Chen} et~al.}{2024}]{Chen2024}
{Chen} Z.,  {Stark} D.~P.,  {Mason} C.,  {Topping} M.~W.,  {Whitler} L.,
  {Tang} M.,  {Endsley} R.,   {Charlot} S.,  2024, \mn@doi [\mnras]
  {10.1093/mnras/stae455}, \href
  {https://ui.adsabs.harvard.edu/abs/2024MNRAS.528.7052C} {528, 7052}

\bibitem[\protect\citeauthoryear{{Choustikov} et~al.,}{{Choustikov}
  et~al.}{2024}]{Choustikov2024}
{Choustikov} N.,  et~al., 2024, \mn@doi [\mnras] {10.1093/mnras/stae1586},
  \href {https://ui.adsabs.harvard.edu/abs/2024MNRAS.532.2463C} {532, 2463}

\bibitem[\protect\citeauthoryear{{Costa}, {Arrigoni Battaia}, {Farina},
  {Keating}, {Rosdahl}  \& {Kimm}}{{Costa} et~al.}{2022}]{Costa2022}
{Costa} T.,  {Arrigoni Battaia} F.,  {Farina} E.~P.,  {Keating} L.~C.,
  {Rosdahl} J.,   {Kimm} T.,  2022, \mn@doi [\mnras] {10.1093/mnras/stac2432},
  \href {https://ui.adsabs.harvard.edu/abs/2022MNRAS.517.1767C} {517, 1767}

\bibitem[\protect\citeauthoryear{{D'Eugenio} et~al.,}{{D'Eugenio}
  et~al.}{2024}]{DEugenio2024}
{D'Eugenio} F.,  et~al., 2024, \mn@doi [arXiv e-prints]
  {10.48550/arXiv.2404.06531}, \href
  {https://ui.adsabs.harvard.edu/abs/2024arXiv240406531D} {p. arXiv:2404.06531}

\bibitem[\protect\citeauthoryear{{Dickinson} et~al.,}{{Dickinson}
  et~al.}{2024}]{Dickinson2024}
{Dickinson} M.,  et~al., 2024, {The CANDELS-Area Prism Epoch of Reionization
  Survey (CAPERS)}, JWST Proposal. Cycle 3, ID. \#6368

\bibitem[\protect\citeauthoryear{{Dijkstra}}{{Dijkstra}}{2014}]{Dijkstra2014}
{Dijkstra} M.,  2014, \mn@doi [\pasa] {10.1017/pasa.2014.33}, \href
  {https://ui.adsabs.harvard.edu/abs/2014PASA...31...40D} {31, e040}

\bibitem[\protect\citeauthoryear{{Dijkstra} \& {Loeb}}{{Dijkstra} \&
  {Loeb}}{2008}]{Dijkstra2008}
{Dijkstra} M.,  {Loeb} A.,  2008, \mn@doi [\mnras]
  {10.1111/j.1365-2966.2008.13920.x}, \href
  {https://ui.adsabs.harvard.edu/abs/2008MNRAS.391..457D} {391, 457}

\bibitem[\protect\citeauthoryear{{Eldridge}, {Stanway}, {Xiao}, {McClelland},
  {Taylor}, {Ng}, {Greis}  \& {Bray}}{{Eldridge} et~al.}{2017}]{bpassv211}
{Eldridge} J.~J.,  {Stanway} E.~R.,  {Xiao} L.,  {McClelland} L.~A.~S.,
  {Taylor} G.,  {Ng} M.,  {Greis} S.~M.~L.,   {Bray} J.~C.,  2017, \mn@doi
  [\pasa] {10.1017/pasa.2017.51}, \href
  {https://ui.adsabs.harvard.edu/abs/2017PASA...34...58E} {34, e058}

\bibitem[\protect\citeauthoryear{Ferland, Korista, Verner, Ferguson, Kingdon
  \& Verner}{Ferland et~al.}{1998}]{ferland1998cloudy}
Ferland G.,  Korista K.,  Verner D.,  Ferguson J.,  Kingdon J.,   Verner E.,
  1998, Publications of the Astronomical Society of the Pacific, 110, 761

\bibitem[\protect\citeauthoryear{{Fontana} et~al.,}{{Fontana}
  et~al.}{2010}]{Fontana2010}
{Fontana} A.,  et~al., 2010, \mn@doi [\apjl] {10.1088/2041-8205/725/2/L205},
  \href {https://ui.adsabs.harvard.edu/abs/2010ApJ...725L.205F} {725, L205}

\bibitem[\protect\citeauthoryear{{Gardner} et~al.,}{{Gardner}
  et~al.}{2023}]{Gardner2023}
{Gardner} J.~P.,  et~al., 2023, \mn@doi [\pasp] {10.1088/1538-3873/acd1b5},
  \href {https://ui.adsabs.harvard.edu/abs/2023PASP..135f8001G} {135, 068001}

\bibitem[\protect\citeauthoryear{{Garel}, {Blaizot}, {Rosdahl},
  {Michel-Dansac}, {Haehnelt}, {Katz}, {Kimm}  \& {Verhamme}}{{Garel}
  et~al.}{2021}]{Garel2021}
{Garel} T.,  {Blaizot} J.,  {Rosdahl} J.,  {Michel-Dansac} L.,  {Haehnelt}
  M.~G.,  {Katz} H.,  {Kimm} T.,   {Verhamme} A.,  2021, \mn@doi [\mnras]
  {10.1093/mnras/stab990}, \href
  {https://ui.adsabs.harvard.edu/abs/2021MNRAS.504.1902G} {504, 1902}

\bibitem[\protect\citeauthoryear{Grimmett, Karakas, Heger, M{\"u}ller  \&
  West}{Grimmett et~al.}{2020}]{grimmett2020chemical}
Grimmett J.,  Karakas A.~I.,  Heger A.,  M{\"u}ller B.,   West C.,  2020,
  Monthly Notices of the Royal Astronomical Society, 496, 4987

\bibitem[\protect\citeauthoryear{{Hamilton}}{{Hamilton}}{1940}]{Hamilton1940}
{Hamilton} D.~R.,  1940, \mn@doi [Physical Review] {10.1103/PhysRev.58.122},
  \href {https://ui.adsabs.harvard.edu/abs/1940PhRv...58..122H} {58, 122}

\bibitem[\protect\citeauthoryear{{Heintz} et~al.,}{{Heintz}
  et~al.}{2024}]{Heintz2024}
{Heintz} K.~E.,  et~al., 2024, \mn@doi [arXiv e-prints]
  {10.48550/arXiv.2404.02211}, \href
  {https://ui.adsabs.harvard.edu/abs/2024arXiv240402211H} {p. arXiv:2404.02211}

\bibitem[\protect\citeauthoryear{{Henyey} \& {Greenstein}}{{Henyey} \&
  {Greenstein}}{1941}]{Henyey1941}
{Henyey} L.~G.,  {Greenstein} J.~L.,  1941, \mn@doi [\apj] {10.1086/144246},
  \href {https://ui.adsabs.harvard.edu/abs/1941ApJ....93...70H} {93, 70}

\bibitem[\protect\citeauthoryear{{Hoag} et~al.,}{{Hoag}
  et~al.}{2019}]{Hoag2019}
{Hoag} A.,  et~al., 2019, \mn@doi [\mnras] {10.1093/mnras/stz1768}, \href
  {https://ui.adsabs.harvard.edu/abs/2019MNRAS.488..706H} {488, 706}

\bibitem[\protect\citeauthoryear{{Hui} \& {Gnedin}}{{Hui} \&
  {Gnedin}}{1997}]{Hui1997}
{Hui} L.,  {Gnedin} N.~Y.,  1997, \mn@doi [\mnras] {10.1093/mnras/292.1.27},
  \href {https://ui.adsabs.harvard.edu/abs/1997MNRAS.292...27H} {292, 27}

\bibitem[\protect\citeauthoryear{Hunter}{Hunter}{2007}]{hunter2007matplotlib}
Hunter J.~D.,  2007, Computing in science \& engineering, 9, 90

\bibitem[\protect\citeauthoryear{{Jakobsen} et~al.,}{{Jakobsen}
  et~al.}{2022}]{Jakobsen2022}
{Jakobsen} P.,  et~al., 2022, \mn@doi [\aap] {10.1051/0004-6361/202142663},
  \href {https://ui.adsabs.harvard.edu/abs/2022A&A...661A..80J} {661, A80}

\bibitem[\protect\citeauthoryear{{Jiang} et~al.,}{{Jiang}
  et~al.}{2023}]{Jiang2023}
{Jiang} H.,  et~al., 2023, \mn@doi [arXiv e-prints]
  {10.48550/arXiv.2312.04151}, \href
  {https://ui.adsabs.harvard.edu/abs/2023arXiv231204151J} {p. arXiv:2312.04151}

\bibitem[\protect\citeauthoryear{{Jones} et~al.,}{{Jones}
  et~al.}{2024}]{Jones2024}
{Jones} G.~C.,  et~al., 2024, \mn@doi [\aap] {10.1051/0004-6361/202347099},
  \href {https://ui.adsabs.harvard.edu/abs/2024A&A...683A.238J} {683, A238}

\bibitem[\protect\citeauthoryear{Jones, Oliphant, Peterson  et~al.}{Jones
  et~al.}{01  }]{scipy}
Jones E.,  Oliphant T.,  Peterson P.,   et~al., 2001--, {SciPy}: Open source
  scientific tools for {Python}, \url {http://www.scipy.org/}

\bibitem[\protect\citeauthoryear{{Jung} et~al.,}{{Jung}
  et~al.}{2020}]{Jung2020}
{Jung} I.,  et~al., 2020, \mn@doi [\apj] {10.3847/1538-4357/abbd44}, \href
  {https://ui.adsabs.harvard.edu/abs/2020ApJ...904..144J} {904, 144}

\bibitem[\protect\citeauthoryear{{Jung} et~al.,}{{Jung}
  et~al.}{2022}]{Jung2022}
{Jung} I.,  et~al., 2022, \mn@doi [arXiv e-prints] {10.48550/arXiv.2212.09850},
  \href {https://ui.adsabs.harvard.edu/abs/2022arXiv221209850J} {p.
  arXiv:2212.09850}

\bibitem[\protect\citeauthoryear{{Jung} et~al.,}{{Jung}
  et~al.}{2023}]{Jung2023}
{Jung} I.,  et~al., 2023, \mn@doi [arXiv e-prints] {10.48550/arXiv.2304.05385},
  \href {https://ui.adsabs.harvard.edu/abs/2023arXiv230405385J} {p.
  arXiv:2304.05385}

\bibitem[\protect\citeauthoryear{{Katz} et~al.,}{{Katz}
  et~al.}{2022}]{2022KatzMgII}
{Katz} H.,  et~al., 2022, \mn@doi [\mnras] {10.1093/mnras/stac1437}, \href
  {https://ui.adsabs.harvard.edu/abs/2022MNRAS.515.4265K} {515, 4265}

\bibitem[\protect\citeauthoryear{{Khusanova} et~al.,}{{Khusanova}
  et~al.}{2020}]{Khusanova2020}
{Khusanova} Y.,  et~al., 2020, \mn@doi [\aap] {10.1051/0004-6361/201935400},
  \href {https://ui.adsabs.harvard.edu/abs/2020A&A...634A..97K} {634, A97}

\bibitem[\protect\citeauthoryear{Kimm \& Cen}{Kimm \&
  Cen}{2014}]{kimm2014escape}
Kimm T.,  Cen R.,  2014, The Astrophysical Journal, 788, 121

\bibitem[\protect\citeauthoryear{Kimm, Cen, Devriendt, Dubois  \& Slyz}{Kimm
  et~al.}{2015}]{kimm2015towards}
Kimm T.,  Cen R.,  Devriendt J.,  Dubois Y.,   Slyz A.,  2015, Monthly Notices
  of the Royal Astronomical Society, 451, 2900

\bibitem[\protect\citeauthoryear{Kretschmer \& Teyssier}{Kretschmer \&
  Teyssier}{2020}]{kretschmer2020forming}
Kretschmer M.,  Teyssier R.,  2020, Monthly Notices of the Royal Astronomical
  Society, 492, 1385

\bibitem[\protect\citeauthoryear{Larson et~al.,}{Larson
  et~al.}{2023}]{Larson_2023}
Larson R.~L.,  et~al., 2023, \mn@doi [The Astrophysical Journal Letters]
  {10.3847/2041-8213/ace619}, 953, L29

\bibitem[\protect\citeauthoryear{{Lemaux} et~al.,}{{Lemaux}
  et~al.}{2021}]{Lemaux2021}
{Lemaux} B.~C.,  et~al., 2021, \mn@doi [\mnras] {10.1093/mnras/stab924}, \href
  {https://ui.adsabs.harvard.edu/abs/2021MNRAS.504.3662L} {504, 3662}

\bibitem[\protect\citeauthoryear{{Li} \& {Draine}}{{Li} \&
  {Draine}}{2001}]{Li2001}
{Li} A.,  {Draine} B.~T.,  2001, \mn@doi [\apj] {10.1086/323147}, \href
  {https://ui.adsabs.harvard.edu/abs/2001ApJ...554..778L} {554, 778}

\bibitem[\protect\citeauthoryear{{Marchi} et~al.,}{{Marchi}
  et~al.}{2017}]{Marchi2017}
{Marchi} F.,  et~al., 2017, \mn@doi [\aap] {10.1051/0004-6361/201630054}, \href
  {https://ui.adsabs.harvard.edu/abs/2017A&A...601A..73M} {601, A73}

\bibitem[\protect\citeauthoryear{Michel-Dansac, Blaizot, Garel, Verhamme, Kimm
  \& Trebitsch}{Michel-Dansac et~al.}{2020}]{Michel_Dansac_2020}
Michel-Dansac L.,  Blaizot J.,  Garel T.,  Verhamme A.,  Kimm T.,   Trebitsch
  M.,  2020, \mn@doi [Astronomy &amp; Astrophysics]
  {10.1051/0004-6361/201834961}, 635, A154

\bibitem[\protect\citeauthoryear{{Nakane} et~al.,}{{Nakane}
  et~al.}{2024}]{Nakane2024}
{Nakane} M.,  et~al., 2024, \mn@doi [\apj] {10.3847/1538-4357/ad38c2}, \href
  {https://ui.adsabs.harvard.edu/abs/2024ApJ...967...28N} {967, 28}

\bibitem[\protect\citeauthoryear{{Nanayakkara} et~al.,}{{Nanayakkara}
  et~al.}{2024}]{Nanayakkara2024}
{Nanayakkara} T.,  et~al., 2024, \mn@doi [Scientific Reports]
  {10.1038/s41598-024-52585-4}, \href
  {https://ui.adsabs.harvard.edu/abs/2024NatSR..14.3724N} {14, 3724}

\bibitem[\protect\citeauthoryear{{Napolitano} et~al.,}{{Napolitano}
  et~al.}{2023}]{Napolitano2023}
{Napolitano} L.,  et~al., 2023, \mn@doi [\aap] {10.1051/0004-6361/202347026},
  \href {https://ui.adsabs.harvard.edu/abs/2023A&A...677A.138N} {677, A138}

\bibitem[\protect\citeauthoryear{{Napolitano} et~al.,}{{Napolitano}
  et~al.}{2024}]{Napolitano2024}
{Napolitano} L.,  et~al., 2024, \mn@doi [arXiv e-prints]
  {10.48550/arXiv.2402.11220}, \href
  {https://ui.adsabs.harvard.edu/abs/2024arXiv240211220N} {p. arXiv:2402.11220}

\bibitem[\protect\citeauthoryear{{Navarre} et~al.,}{{Navarre}
  et~al.}{2024}]{2024Navarre}
{Navarre} A.,  et~al., 2024, \mn@doi [\apj] {10.3847/1538-4357/ad10ad}, \href
  {https://ui.adsabs.harvard.edu/abs/2024ApJ...962..175N} {962, 175}

\bibitem[\protect\citeauthoryear{{Ning} et~al.,}{{Ning}
  et~al.}{2024}]{Ning2024}
{Ning} Y.,  et~al., 2024, \mn@doi [\apjl] {10.3847/2041-8213/ad292f}, \href
  {https://ui.adsabs.harvard.edu/abs/2024ApJ...963L..38N} {963, L38}

\bibitem[\protect\citeauthoryear{{Ouchi} et~al.,}{{Ouchi}
  et~al.}{2008}]{Ouchi2008}
{Ouchi} M.,  et~al., 2008, \mn@doi [\apjs] {10.1086/527673}, \href
  {https://ui.adsabs.harvard.edu/abs/2008ApJS..176..301O} {176, 301}

\bibitem[\protect\citeauthoryear{{Ouchi}, {Ono}  \& {Shibuya}}{{Ouchi}
  et~al.}{2020}]{Ouchi2020}
{Ouchi} M.,  {Ono} Y.,   {Shibuya} T.,  2020, \mn@doi [\araa]
  {10.1146/annurev-astro-032620-021859}, \href
  {https://ui.adsabs.harvard.edu/abs/2020ARA&A..58..617O} {58, 617}

\bibitem[\protect\citeauthoryear{{Partridge} \& {Peebles}}{{Partridge} \&
  {Peebles}}{1967}]{1967partridge}
{Partridge} R.~B.,  {Peebles} P.~J.~E.,  1967, \mn@doi [\apj] {10.1086/149161},
  \href {https://ui.adsabs.harvard.edu/abs/1967ApJ...148..377P} {148, 377}

\bibitem[\protect\citeauthoryear{{Pentericci} et~al.,}{{Pentericci}
  et~al.}{2011}]{Pentericci2011}
{Pentericci} L.,  et~al., 2011, \mn@doi [\apj] {10.1088/0004-637X/743/2/132},
  \href {https://ui.adsabs.harvard.edu/abs/2011ApJ...743..132P} {743, 132}

\bibitem[\protect\citeauthoryear{{Pentericci} et~al.,}{{Pentericci}
  et~al.}{2018}]{Pentericci2018}
{Pentericci} L.,  et~al., 2018, \mn@doi [\aap] {10.1051/0004-6361/201833047},
  \href {http://adsabs.harvard.edu/abs/2018A%26A...616A.174P} {616, A174}

\bibitem[\protect\citeauthoryear{{Planck Collaboration} et~al.,}{{Planck
  Collaboration} et~al.}{2016}]{Planck2016}
{Planck Collaboration} et~al., 2016, \mn@doi [\aap]
  {10.1051/0004-6361/201525830}, \href
  {https://ui.adsabs.harvard.edu/abs/2016A&A...594A..13P} {594, A13}

\bibitem[\protect\citeauthoryear{{Ribeiro} et~al.,}{{Ribeiro}
  et~al.}{2020}]{Ribeiro2020}
{Ribeiro} B.,  et~al., 2020, \mn@doi [arXiv e-prints]
  {10.48550/arXiv.2007.01322}, \href
  {https://ui.adsabs.harvard.edu/abs/2020arXiv200701322R} {p. arXiv:2007.01322}

\bibitem[\protect\citeauthoryear{{Richard} et~al.,}{{Richard}
  et~al.}{2021}]{Richard2021}
{Richard} J.,  et~al., 2021, \mn@doi [\aap] {10.1051/0004-6361/202039462},
  \href {https://ui.adsabs.harvard.edu/abs/2021A&A...646A..83R} {646, A83}

\bibitem[\protect\citeauthoryear{{Rosdahl} \& {Teyssier}}{{Rosdahl} \&
  {Teyssier}}{2015a}]{rosdahl2015}
{Rosdahl} J.,  {Teyssier} R.,  2015a, \mn@doi [\mnras] {10.1093/mnras/stv567},
  \href {https://ui.adsabs.harvard.edu/abs/2015MNRAS.449.4380R} {449, 4380}

\bibitem[\protect\citeauthoryear{Rosdahl \& Teyssier}{Rosdahl \&
  Teyssier}{2015b}]{rosdahl2015scheme}
Rosdahl J.,  Teyssier R.,  2015b, Monthly Notices of the Royal Astronomical
  Society, 449, 4380

\bibitem[\protect\citeauthoryear{{Rosdahl}, {Blaizot}, {Aubert}, {Stranex}  \&
  {Teyssier}}{{Rosdahl} et~al.}{2013a}]{rosdahl2013}
{Rosdahl} J.,  {Blaizot} J.,  {Aubert} D.,  {Stranex} T.,   {Teyssier} R.,
  2013a, \mn@doi [\mnras] {10.1093/mnras/stt1722}, \href
  {https://ui.adsabs.harvard.edu/abs/2013MNRAS.436.2188R} {436, 2188}

\bibitem[\protect\citeauthoryear{Rosdahl, Blaizot, Aubert, Stranex  \&
  Teyssier}{Rosdahl et~al.}{2013b}]{rosdahl2013ramses}
Rosdahl J.,  Blaizot J.,  Aubert D.,  Stranex T.,   Teyssier R.,  2013b,
  Monthly Notices of the Royal Astronomical Society, 436, 2188

\bibitem[\protect\citeauthoryear{Rosen \& Bregman}{Rosen \&
  Bregman}{1995}]{rosen1995global}
Rosen A.,  Bregman J.~N.,  1995, The Astrophysical Journal, 440, 634

\bibitem[\protect\citeauthoryear{{Saxena} et~al.,}{{Saxena}
  et~al.}{2024}]{Saxena2024}
{Saxena} A.,  et~al., 2024, \mn@doi [\aap] {10.1051/0004-6361/202347132}, \href
  {https://ui.adsabs.harvard.edu/abs/2024A&A...684A..84S} {684, A84}

\bibitem[\protect\citeauthoryear{{Schenker}, {Ellis}, {Konidaris}  \&
  {Stark}}{{Schenker} et~al.}{2014}]{Schenker2014}
{Schenker} M.~A.,  {Ellis} R.~S.,  {Konidaris} N.~P.,   {Stark} D.~P.,  2014,
  \mn@doi [\apj] {10.1088/0004-637X/795/1/20}, \href
  {https://ui.adsabs.harvard.edu/abs/2014ApJ...795...20S} {795, 20}

\bibitem[\protect\citeauthoryear{{Schmidt} et~al.,}{{Schmidt}
  et~al.}{2021}]{Schmidt2021}
{Schmidt} K.~B.,  et~al., 2021, \mn@doi [\aap] {10.1051/0004-6361/202140876},
  \href {https://ui.adsabs.harvard.edu/abs/2021A&A...654A..80S} {654, A80}

\bibitem[\protect\citeauthoryear{{Smith}, {Safranek-Shrader}, {Bromm}  \&
  {Milosavljevi{\'c}}}{{Smith} et~al.}{2015}]{Smith2015}
{Smith} A.,  {Safranek-Shrader} C.,  {Bromm} V.,   {Milosavljevi{\'c}} M.,
  2015, \mn@doi [\mnras] {10.1093/mnras/stv565}, \href
  {https://ui.adsabs.harvard.edu/abs/2015MNRAS.449.4336S} {449, 4336}

\bibitem[\protect\citeauthoryear{{Smith}, {Kannan}, {Garaldi}, {Vogelsberger},
  {Pakmor}, {Springel}  \& {Hernquist}}{{Smith} et~al.}{2022}]{Smith2022}
{Smith} A.,  {Kannan} R.,  {Garaldi} E.,  {Vogelsberger} M.,  {Pakmor} R.,
  {Springel} V.,   {Hernquist} L.,  2022, \mn@doi [\mnras]
  {10.1093/mnras/stac713}, \href
  {https://ui.adsabs.harvard.edu/abs/2022MNRAS.512.3243S} {512, 3243}

\bibitem[\protect\citeauthoryear{{Stanway} \& {Eldridge}}{{Stanway} \&
  {Eldridge}}{2018}]{bpassv221}
{Stanway} E.~R.,  {Eldridge} J.~J.,  2018, \mn@doi [\mnras]
  {10.1093/mnras/sty1353}, \href
  {https://ui.adsabs.harvard.edu/abs/2018MNRAS.479...75S} {479, 75}

\bibitem[\protect\citeauthoryear{{Stark}, {Ellis}, {Chiu}, {Ouchi}  \&
  {Bunker}}{{Stark} et~al.}{2010}]{Stark2010}
{Stark} D.~P.,  {Ellis} R.~S.,  {Chiu} K.,  {Ouchi} M.,   {Bunker} A.,  2010,
  \mn@doi [\mnras] {10.1111/j.1365-2966.2010.17227.x}, \href
  {https://ui.adsabs.harvard.edu/abs/2010MNRAS.408.1628S} {408, 1628}

\bibitem[\protect\citeauthoryear{{Stark}, {Ellis}  \& {Ouchi}}{{Stark}
  et~al.}{2011}]{Stark2011}
{Stark} D.~P.,  {Ellis} R.~S.,   {Ouchi} M.,  2011, \mn@doi [\apjl]
  {10.1088/2041-8205/728/1/L2}, \href
  {https://ui.adsabs.harvard.edu/abs/2011ApJ...728L...2S} {728, L2}

\bibitem[\protect\citeauthoryear{Sukhbold, Ertl, Woosley, Brown  \&
  Janka}{Sukhbold et~al.}{2016}]{sukhbold2016core}
Sukhbold T.,  Ertl T.,  Woosley S.,  Brown J.~M.,   Janka H.-T.,  2016, The
  Astrophysical Journal, 821, 38

\bibitem[\protect\citeauthoryear{{Tang} et~al.,}{{Tang}
  et~al.}{2023}]{Tang2023}
{Tang} M.,  et~al., 2023, \mn@doi [\mnras] {10.1093/mnras/stad2763}, \href
  {https://ui.adsabs.harvard.edu/abs/2023MNRAS.526.1657T} {526, 1657}

\bibitem[\protect\citeauthoryear{{Tang}, {Stark}, {Topping}, {Mason}  \&
  {Ellis}}{{Tang} et~al.}{2024}]{Tang2024}
{Tang} M.,  {Stark} D.~P.,  {Topping} M.~W.,  {Mason} C.,   {Ellis} R.~S.,
  2024, \mn@doi [arXiv e-prints] {10.48550/arXiv.2408.01507}, \href
  {https://ui.adsabs.harvard.edu/abs/2024arXiv240801507T} {p. arXiv:2408.01507}

\bibitem[\protect\citeauthoryear{Van Der~Walt, Colbert  \& Varoquaux}{Van
  Der~Walt et~al.}{2011}]{van2011numpy}
Van Der~Walt S.,  Colbert S.~C.,   Varoquaux G.,  2011, Computing in science \&
  engineering, 13, 22

\bibitem[\protect\citeauthoryear{{Verhamme}, {Dubois}, {Blaizot}, {Garel},
  {Bacon}, {Devriendt}, {Guiderdoni}  \& {Slyz}}{{Verhamme}
  et~al.}{2012}]{Verhamme2012}
{Verhamme} A.,  {Dubois} Y.,  {Blaizot} J.,  {Garel} T.,  {Bacon} R.,
  {Devriendt} J.,  {Guiderdoni} B.,   {Slyz} A.,  2012, \mn@doi [\aap]
  {10.1051/0004-6361/201218783}, \href
  {https://ui.adsabs.harvard.edu/abs/2012A&A...546A.111V} {546, A111}

\bibitem[\protect\citeauthoryear{{Verhamme}, {Orlitov{\'a}}, {Schaerer}  \&
  {Hayes}}{{Verhamme} et~al.}{2015}]{Verhamme2015}
{Verhamme} A.,  {Orlitov{\'a}} I.,  {Schaerer} D.,   {Hayes} M.,  2015, \mn@doi
  [\aap] {10.1051/0004-6361/201423978}, \href
  {https://ui.adsabs.harvard.edu/abs/2015A&A...578A...7V} {578, A7}

\bibitem[\protect\citeauthoryear{{Whitney}}{{Whitney}}{2011}]{Whitney2011}
{Whitney} B.~A.,  2011, \mn@doi [Bulletin of the Astronomical Society of India]
  {10.48550/arXiv.1104.4990}, \href
  {https://ui.adsabs.harvard.edu/abs/2011BASI...39..101W} {39, 101}

\bibitem[\protect\citeauthoryear{{Wisotzki} et~al.,}{{Wisotzki}
  et~al.}{2018}]{Wisotzki2018}
{Wisotzki} L.,  et~al., 2018, \mn@doi [\nat] {10.1038/s41586-018-0564-6}, \href
  {https://ui.adsabs.harvard.edu/abs/2018Natur.562..229W} {562, 229}

\bibitem[\protect\citeauthoryear{{Yuan}, {Martin-Alvarez}, {Haehnelt}, {Garel}
  \& {Sijacki}}{{Yuan} et~al.}{2024}]{Yuxuan2024}
{Yuan} Y.,  {Martin-Alvarez} S.,  {Haehnelt} M.~G.,  {Garel} T.,   {Sijacki}
  D.,  2024, \mn@doi [\mnras] {10.1093/mnras/stae1606}, \href
  {https://ui.adsabs.harvard.edu/abs/2024MNRAS.532.3643Y} {532, 3643}

\bibitem[\protect\citeauthoryear{{Zheng} \& {Miralda-Escud{\'e}}}{{Zheng} \&
  {Miralda-Escud{\'e}}}{2002}]{Zheng2002}
{Zheng} Z.,  {Miralda-Escud{\'e}} J.,  2002, \mn@doi [\apj] {10.1086/342400},
  \href {https://ui.adsabs.harvard.edu/abs/2002ApJ...578...33Z} {578, 33}

\bibitem[\protect\citeauthoryear{{Zheng}, {Malhotra}, {Rhoads}, {Finkelstein},
  {Wang}, {Jiang}  \& {Cai}}{{Zheng} et~al.}{2016}]{Zheng2016}
{Zheng} Z.-Y.,  {Malhotra} S.,  {Rhoads} J.~E.,  {Finkelstein} S.~L.,  {Wang}
  J.-X.,  {Jiang} C.-Y.,   {Cai} Z.,  2016, \mn@doi [\apjs]
  {10.3847/0067-0049/226/2/23}, \href
  {https://ui.adsabs.harvard.edu/abs/2016ApJS..226...23Z} {226, 23}

\makeatother
\end{thebibliography}





\bsp	
\label{lastpage}
\end{document}